\pdfoutput=1
\documentclass[sigconf,natbib=true]{acmart}
\usepackage{soul}
\usepackage{url}
\usepackage[utf8]{inputenc}
\usepackage{graphicx}
\usepackage{amsmath}
\usepackage{amsthm}
\usepackage{booktabs}
\usepackage{algorithm}
\usepackage{algorithmic}
\usepackage{amsfonts}
\usepackage{subcaption}
\usepackage[nolist,nohyperlinks]{acronym}
\usepackage{hyperref}
\usepackage{lipsum} 
\usepackage{array}
\usepackage{varwidth}

\newcommand{\up}[1]{\textsuperscript{#1}}

\AtBeginDocument{%
  \providecommand\BibTeX{{%
    \normalfont B\kern-0.5em{\scshape i\kern-0.25em b}\kern-0.8em\TeX}}}

\copyrightyear{2023}
\acmYear{2023}
\setcopyright{rightsretained}
\acmConference[SIGIR '23]{Proceedings of the 46th International ACM SIGIR Conference on Research and Development in Information Retrieval}{July 23--27, 2023}{Taipei, Taiwan}
\acmBooktitle{Proceedings of the 46th International ACM SIGIR Conference on Research and Development in Information Retrieval (SIGIR '23), July 23--27, 2023, Taipei, Taiwan}
\acmPrice{15.00}
\acmDOI{10.1145/3539618.3591628}
\acmISBN{978-1-4503-9408-6/23/07}

\begin{document}

\title{A Scalable Framework for Automatic Playlist Continuation on~Music~Streaming~Services}

\author{Walid Bendada}
\affiliation{
  \institution{Deezer Research}
  \institution{LAMSADE, Université Paris Dauphine, PSL}
  \city{}
  \country{}
}
\email{research@deezer.com}

\author{Guillaume Salha-Galvan}
\affiliation{
  \institution{Deezer Research}
  \city{}
  \country{}
}

\author{Thomas Bouabça}
\affiliation{
  \institution{Deezer Research}
  \city{}
  \country{}
}

\author{Tristan Cazenave}
\affiliation{
  \institution{LAMSADE, Université Paris Dauphine, PSL}
  \city{}
  \country{}
}

\renewcommand{\shortauthors}{W. Bendada, et al.}

\begin{abstract}

Music streaming services often aim to recommend songs for users to extend the playlists they have created on these services. However, extending playlists while preserving their musical characteristics and matching user preferences remains a challenging task, commonly referred to as \textit{Automatic Playlist Continuation}~(APC). Besides, while these services often need to select the best songs to recommend in real-time and among large catalogs with millions of candidates, recent research on APC mainly focused on models with few scalability guarantees and evaluated on relatively small datasets.
In~this~paper, we introduce a general framework to build scalable yet effective APC models for large-scale applications. Based on a represent-then-aggregate strategy, it ensures scalability by design while remaining flexible enough to incorporate a wide range of representation learning and sequence modeling techniques, e.g., based on Transformers. We demonstrate the relevance of this framework through in-depth experimental validation on Spotify's Million Playlist Dataset~(MPD), the largest public dataset for APC.
We also describe how, in 2022, we successfully leveraged this framework to improve APC in production on Deezer. We report results from a large-scale online A/B test on this service, emphasizing the practical impact of our approach in such a~real-world~application.
\end{abstract}





\maketitle

\section{Introduction}
\label{sec:intro}

Recommender systems are becoming increasingly important for music streaming services such as Apple Music, Deezer, or~Spotify \cite{Briand2021AApps,jacobson2016music,Schedl2018CurrentResearch}. While these services provide access to ever-growing musical catalogs, their recommender systems prevent information overload problems by identifying the most relevant content to showcase to each user~\cite{bobadilla2013recommender,schedl2019deep}. Recommender systems also enable users to discover new songs, albums, or artists they may like~\cite{jacobson2016music,Schedl2018CurrentResearch}. Overall, they are widely regarded as effective tools to improve the user experience and engagement on these services~\cite{bobadilla2013recommender,zhang2019deep,mu2018survey,bontempelli2022flow}.

In particular, music streaming services often recommend songs for users to continue the personal \textit{playlists} they have created on these services. Broadly defined as ordered sequences of songs intended to be listened to together, playlists often comply with specific music genres, moods, cultural themes, or activities~\cite{Bendada2020CarouselBandits,bontempelli2022flow,Zamani2019AnContinuation}. Automatically extending them while preserving their musical characteristics and matching user preferences remains a challenging task, commonly referred to as \textit{Automatic Playlist Continuation}~(APC) \cite{Schedl2018CurrentResearch,Chen2018RecSysContinuation,Zamani2019AnContinuation,Volkovs2018Two-stageScale}. Although there have been attempts to tackle APC predating music streaming services (see, e.g., the survey of Bonnin et al.~\cite{Bonnin2014AutomatedExperiments}), the advent of these services has driven music listeners towards more and more playlist consumption~\cite{jakobsen2016playlists}, and the need for viable APC solutions has been growing ever since. APC is now considered one of the ``\textit{most pressing current challenges in music recommender systems research}'' by Schedl et al.~\cite{Schedl2018CurrentResearch}.


As APC is closely related to sequence-based and session-based recommendation \cite{Wang2019Sequential,Dang2020ASystems}, recent research on this problem extensively focused on leveraging models already successful in other sequence modeling tasks, notably language modeling, for recommendation purposes. Drawing on the analogy that exists between words within sentences on the one hand, and songs within playlists on the other hand, researchers have proposed various effective APC models based on word2vec \cite{Vasile2016Meta-Prod2VecRecommendation,Wang2016LearningRecommendation}, recurrent neural networks~\cite{Jing2017NeuralRecommender,Donkers2017SequentialRecommendations}, convolutional neural networks~\cite{Yuan2019ARecommendation}, and attention mechanisms~\cite{Guo2019StreamingRecommendation}. However, as we will detail in Section~\ref{section:related}, these promising studies often overlooked scalability concerns and evaluated their models on relatively small subsets of~real-world~datasets.

Yet, scalability is essential for industrial applications on music streaming services, which must select the best songs to recommend for APC among large catalogs with tens of millions of candidates. 
Besides the number of songs to handle when training the APC model, scalability of inference, while being sometimes neglected, is
also crucial. As we will further elaborate in Sections~\ref{section:continuation}~to~\ref{section:online_experiments}, music streaming services must usually perform APC online, repeatedly, and in real-time. 
Indeed, users regularly create new playlists and modify the existing ones. In return, they expect the system to provide updated recommendations, with~minimal~delay.

 In summary, there is a discrepancy between the complexity and evaluation procedure of APC models proposed in the recent scientific literature, and the scalability requirements associated with industrial applications. As an illustration of this discrepancy, we stress that modern sequence modeling techniques were seldom used during the RecSys~2018 APC Challenge~\cite{jannach2020deep,Zamani2019AnContinuation}. Aiming to foster research on large-scale APC, this open challenge was evaluated on Spotify's Million Playlist Dataset (MPD), which, to this day, remains the largest public dataset for APC~\cite{Chen2018RecSysContinuation}. To provide APC recommendations on this dataset, many teams favored relatively simpler methods during the challenge (see~our~review~in~Section~\ref{section:related}) and, as of today, leveraging the modern APC models discussed above in such a large-scale setting still poses~scalability~challenges.


In this paper, we propose a general framework to overcome these challenges and remedy this observed discrepancy. While remaining versatile enough to incorporate a wide range of modern methods, our solution formally characterizes the requirements expected from suitable APC solutions for large-scale industrial applications. More precisely, our contributions in this paper are listed as follows:

\begin{itemize}
    \item We introduce a principled framework to build scalable yet effective APC models. Based on a \textit{represent-then-aggregate} strategy, it permits incorporating a wide range of complex APC models into large-scale systems suitable for industrial applications. Our framework systematically decomposes these models into a part handling song representation learning, and a part dedicated to playlist-level~sequence~modeling.
    \item We illustrate the possibilities induced by our framework, showing how one can design scalable APC models benefiting from the most popular architectures for sequence modeling, e.g., recurrent neural networks~\cite{Hidasi2016Session-basedNetworks} or Transformers~\cite{Vaswani2017AttentionNeed}, combined with complex song representation learning models, e.g., neural architectures processing metadata~\cite{rendle2010factorization}.
    \item We demonstrate the empirical relevance of our framework for large-scale APC through in-depth experimental validation on Spotify's MPD dataset, the largest public APC dataset. Along with this paper, we publicly release our source code on GitHub to ensure the reproducibility of our results.
    \item We also describe how, in 2022, we leveraged this framework to improve APC in production on the global music streaming service Deezer. We present results from a large-scale online A/B test on users from this service, emphasizing the practical impact of our framework in such a real-world~application.
\end{itemize}

This article is organized as follows. In Section~\ref{section:related}, we introduce the APC problem more precisely and review previous work. In Section~\ref{section:continuation}, we present our proposed framework to build scalable yet effective APC models, providing a detailed overview of its possibilities and limitations. We report our experimental setting and results on the public MPD dataset in Section~\ref{section:offline_experiments}, and describe our online A/B test on the Deezer service in Section~\ref{section:online_experiments}. 
Finally, we conclude~in~Section~\ref{section:conclusion}.

\section{Preliminaries}
\label{section:related}

We begin this section by precisely defining the APC problem we aim to solve. We subsequently review the relevant related work.

\subsection{Problem Formulation}
\label{section:s21}

\subsubsection{Notation}
\begin{figure}[t]
  \centering
  \includegraphics[width=0.68\linewidth]{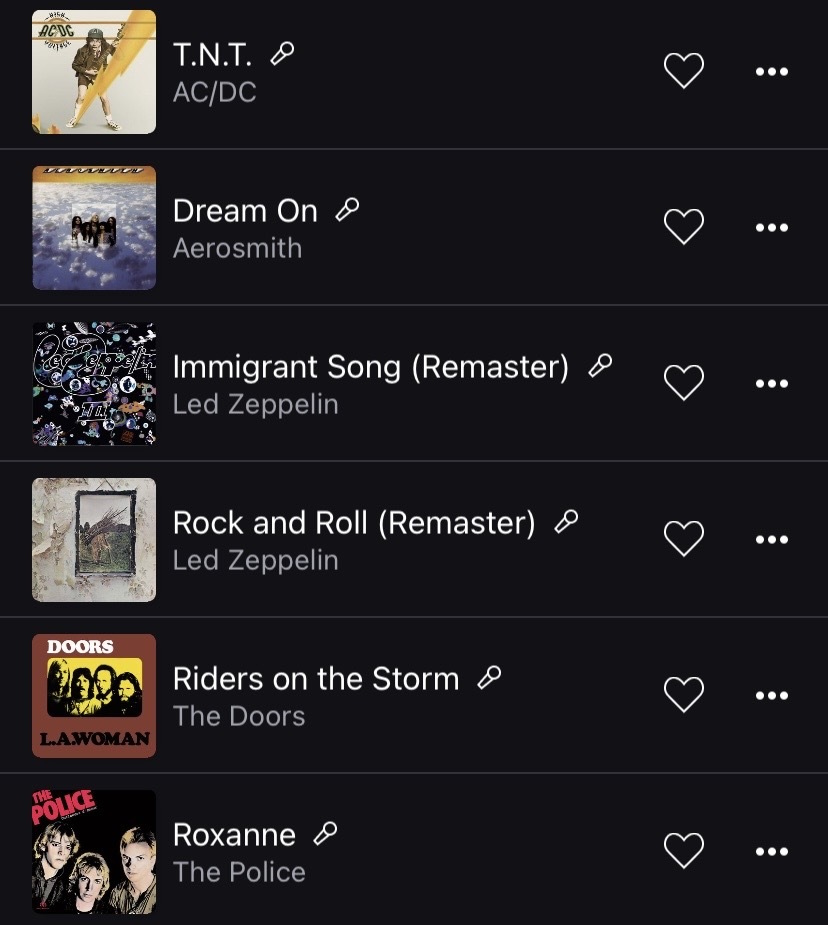}
  \caption{An example of a user playlist on Deezer.} 
\label{figure:playlist1}
\end{figure}

This paper considers a catalog $\mathcal{S}~=~\{s_1, ..., s_N\}$ of $N \in \mathbb{N}^*$ songs available on a music streaming service. Users from this service can create playlists containing these songs, as in Figure~\ref{figure:playlist1}. They can update their existing playlists at any time, by adding, removing, or reordering songs. 
We denote by $L \in \mathbb{N}^*$ the maximum length for a playlist, a parameter fixed by the service. We denote by $\mathcal{P}$ the set of all playlists that can be created from $\mathcal{S}$:
\begin{equation}
\mathcal{P} = \bigcup_{l=1}^{L} \mathcal{S}^l.
\end{equation}
Lastly, we associate each song $s \in \mathcal{S}$ with a descriptive tuple $m_s$ of size $M \in \mathbb{N}^*$. This tuple captures metadata information on the song, e.g., the name of the artist or the album it~belongs~to:
\begin{equation}
m_s \in \mathcal{M} = \mathcal{M}_1 \times \mathcal{M}_2 \times \dots \times \mathcal{M}_{M}.
\end{equation}
In this equation, each $\mathcal{M}_i$ denotes the set of possible values for a given metadata type, e.g., the set of possible artists. 

\subsubsection{APC}
\label{s212}
In such a setting, the APC problem consists in extending a playlist $p \in \mathcal{P}$ by adding more songs matching the target characteristics of $p$~\cite{Schedl2018CurrentResearch,Chen2018RecSysContinuation}. Formally, an APC~model~is~a~function:
\begin{equation}
f \colon \mathcal{P} \times \mathcal{S} \to \mathbb{R},
\label{eqAPC}
\end{equation}
associating each $p \in \mathcal{P}$ and $s \in \mathcal{S}$ with a ``relevance'' or ``similarity'' score $f(p,s)$. The higher the score, the more likely $s$ will fit well as a continuation of $p$, according to~$f$. To assess and compare the performance of APC models, the standard methodology~consists~in:
\begin{itemize}
    \item Collecting a set of test playlists, unseen during training.
    \item Then, masking the last songs of these test playlists.
    \item Finally, evaluating the ability of each model to complete these partially masked sequences using song relevance scores, computing metrics such as the ones we will consider in Section~\ref{section:offline_experiments}.
\end{itemize}

\subsection{Related Work}
\label{section:s22}
 In recent years, the APC problem has garnered significant attention from researchers, leading to substantial efforts~to~address~it.

\subsubsection{Collaborative Filtering for APC} 
\label{s221}
Historically, Collaborative Filtering (CF)~\cite{koren2015advances,su2009survey} has been a prevalent approach for APC~\cite{Faggioli2018EfficientTask,Ludewig2018EffectiveRecommendations,Volkovs2018Two-stageScale,Schedl2018CurrentResearch}. As illustrated in Figure~\ref{fig:matrix}, CF methods for APC usually represent playlist datasets as sparse matrices where each row corresponds to a playlist, each column corresponds to a song, and the associated binary value indicates whether the song appears in the playlist. In essence, they aim to infer the relevance of a song within a playlist by leveraging the content~of~similar~playlists.
Some of these CF methods compute inner products between pairs of playlists sparse vectors on one side, pairs of songs sparse vectors on the other side, and use the information of song occurrence within playlists to compute similarities between playlists and songs \cite{Faggioli2018EfficientTask,Volkovs2018Two-stageScale}.
Others assume that the sparse playlist-song occurrence matrix has a low-rank structure and depict it as the product of two rectangular dense matrices using Matrix Factorization (MF) techniques~\cite{koren2009matrix,Ludewig2018EffectiveRecommendations}. The first matrix, of dimension $K \times D$ (with $K$ the number of playlists in the dataset, and some fixed $D \ll \min(K,N)$), represents a playlist by row. The second one, of dimension $D \times N$, represents a song by column. They can be interpreted as $D$-dimensional vectorial representations of playlists and songs in a common ``embedding'' vector space. In this space, one can compute the similarity between any pair, e.g., using inner products \cite{Hu2008CollaborativeYifan}.
As these CF methods usually fail to provide recommendations for songs absent from existing playlists~\cite{rendle2010factorization,gope2017survey}, researchers have also proposed to integrate metadata information into predictions, with the underlying assumption that co-occurrence information of frequent songs should flow to rare songs sharing similar attributes. Notable examples of CF models integrating metadata include Factorization Machines (FM) \cite{rendle2010factorization}. FM represent metadata in the same embedding space as playlists and songs, and add inner products between embedding vectors of various metadata to the original prediction score. They have been successfully used~for~APC~\cite{Ludewig2018EffectiveRecommendations,Yang2018MMCF:Continuation,Rubtsov2018AContinuation,Ferraro2018AutomaticAudio}.

\subsubsection{Sequence Modeling for APC}
\label{s222} 
\begin{figure}
     \centering
      \begin{subfigure}[b]{0.38\linewidth}
             \raggedright
          \includegraphics[width=\linewidth]{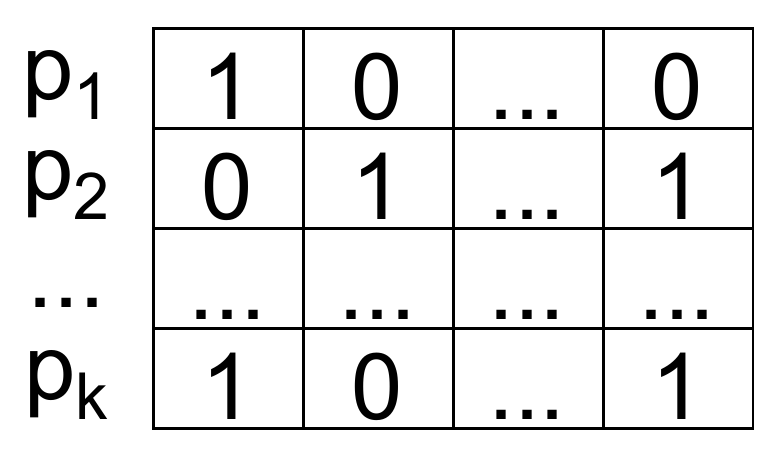}
        \caption{Sparse matrix}
        \label{fig:matrix}
      \end{subfigure}
      \hfill \vline
\begin{subfigure}[b]{0.6\linewidth}
       \raggedleft
          \includegraphics[width=\linewidth]{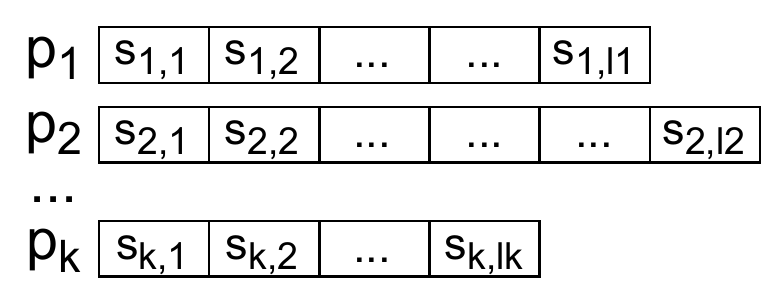}
        \caption{List of sequences}
        \label{fig:sequence}
      \end{subfigure}
      \caption{An APC dataset represented as (\subref{fig:matrix}) a sparse playlist-song occurrence matrix, (\subref{fig:sequence}) a list of uneven~song~sequences.}
      \label{representation}
\end{figure}

Over the past few years, an increasing number of research works on APC have proposed to represent playlist datasets, not as sparse interaction matrices, but as uneven lists of song sequences, as illustrated in Figure~\ref{fig:sequence}. These studies have adapted techniques already successful in other sequence modeling tasks, notably language modeling, to extend the ordered sequences of songs that playlists naturally constitute \cite{Vasile2016Meta-Prod2VecRecommendation,Wang2016LearningRecommendation,Donkers2017SequentialRecommendations,Guo2019StreamingRecommendation,Yuan2019ARecommendation}. More specifically, they have introduced APC models based on word2vec \cite{Vasile2016Meta-Prod2VecRecommendation,Wang2016LearningRecommendation}, recurrent neural networks~\cite{Jing2017NeuralRecommender,Donkers2017SequentialRecommendations}, convolutional neural networks~\cite{Yuan2019ARecommendation}, and attention mechanisms~\cite{Guo2019StreamingRecommendation}. 
In the experimental evaluations of these studies, such sequential methods often outperform the pure CF approaches from Section~\ref{s221}. However, these promising results were obtained on relatively small subsets\footnote{In particular, the range of recommendable songs was often restricted to the most popular ones. Besides scalability concerns, such a restriction also questions the ability of these models to recommend songs from the \textit{long~tail}, a matter of great importance considering that the size of the catalog is one of the specificities of music recommendation \cite{Schedl2018CurrentResearch} and that numerous models tend to be biased towards popular artists~\cite{Kowald2019TheStudy}.} of real-world datasets (see, e.g., the first four lines of Table~\ref{table:datasets}), due to the absence of
larger public real-world datasets (for studies predating the MPD~release~\cite{Chen2018RecSysContinuation}) but also to intrinsic scalability issues related to some of these models~\cite{Zamani2019AnContinuation}. Hence, their practical effectiveness in large-scale applications involving millions of songs and playlists still needed to be fully demonstrated. Regarding the most recent class of sequence modeling neural architectures, i.e.,~Transformers~\cite{Vaswani2017AttentionNeed}, they have been proposed for sequential product recommendation (see, e.g., SASRec and BERT4Rec~\cite{Kang2018Self-AttentiveRecommendation, Sun2019BERT4Rec:Transformer}), but we have not seen examples of usage on large-scale APC. Similarly, nearest neighbors techniques have been proposed for sequence prediction, sometimes outperforming complex neural models \cite{Ludewig2018EvaluationAlgorithms,Ludewig2019PerformanceRecommendation,Ludewig2019EmpiricalAlgorithms}, but were not evaluated on large-scale APC. Finally, while some approaches leverage the multi-modal aspect of APC and others focus on its sequential nature, to our knowledge, few attempts have been made to explicitly consider both components while remaining efficient enough for large-scale~applications.

\begin{table}
\centering
\caption{Public datasets for APC.}
\label{table:datasets}
\resizebox{0.4\textwidth}{!}{
\begin{tabular}{c|ccc}
\toprule
\textbf{Dataset} &  \textbf{Songs} & \textbf{Playlists} & \textbf{Interactions} \\
\midrule
\midrule 
AOTM~\cite{McFee2012HypergraphScholar} & 91~166 & 27~005 & 306~830\\
NOWPLAYING~\cite{Zangerle2014nowplayingTwitter} & 75~169 & 75~169 & 271~177 \\
30MUSIC~\cite{Turrin201530MusicDataset} & 210~633 & 37~333 & 638~933 \\
MELON~\cite{Ferraro2021MelonTagging} & 649~091 &  148~826 & 5~904~718 \\
MPD~\cite{Chen2018RecSysContinuation} & 2~262~292& 1~000~000 & 66~346~428\\

\bottomrule
\end{tabular}
}
\label{playlistdatasets}
\end{table}


\subsubsection{Towards Large-Scale APC}
\label{s223}
To foster research on large-scale APC, the 12\up{th} ACM Conference on Recommender Systems hosted the ``RecSys 2018 APC Challenge'' \cite{Chen2018RecSysContinuation}, evaluated on the Million Playlist Dataset (MPD), which was then publicly released. Composed of one million playlists created by Spotify users, the MPD stands out for its magnitude compared with other public datasets from Table~\ref{table:datasets}, as well as for its sparsity. Indeed, while more than two million songs can be found in the dataset, roughly 60\% of them do not appear more than twice within playlists~\cite{Chen2018RecSysContinuation}. Spotify also provided side information on songs and playlists. In practice, such information often helps overcome the cold start problem induced by data sparsity~\cite{Briand2021AApps,salha2021cold}. In 2019, Zamani~et~al.~\cite{Zamani2019AnContinuation} analyzed key insights from the challenge, explaining that many teams leveraged:
\begin{itemize}
\item Ensemble architectures: several teams combined several APC models of the form $f \colon \mathcal{P} \times \mathcal{S} \to \mathbb{R}$, each of them providing different ``candidate'' songs to extend playlists.
\item Or, two-stage architectures: a first simple model rapidly scored all songs from $\mathcal{S}$, retrieving a subset of ``candidate'' songs, several orders of magnitude smaller than the original song set. Then, some more sophisticated model(s) re-ranked candidate songs to improve APC recommendations. The performance of such two-stage strategies was highly dependent on the quality of candidate~retrieval~models~\cite{Zamani2019AnContinuation}.
\end{itemize} 
Overall, the modern sequence modeling neural networks from Section~\ref{s222} were seldom used for candidate retrieval in these systems, either because they were used in a way that did not scale to millions of songs, or led to underperforming results with respect to alternatives~\cite{Zamani2019AnContinuation}.
When considering both candidate selection models of ensemble/two-stage architectures, as well as single-stage architectures directly scoring all songs, most approaches leveraged simpler and faster non-parametric nearest neighbor models, applied to various CF-based song/playlist embedding representations learned using models from Section~\ref{s221}. 
In summary, most teams favored simpler and more
scalable methods than the ones presented in Section~\ref{s222} despite their promising performances on smaller datasets. 
We believe this challenge highlighted the current discrepancy between the complexity and evaluation procedure of APC models proposed in the recent scientific literature, and the simpler ones successfully used for large-scale applications. In particular, in the context of a global questioning of the ability of deep learning to actually improve recommender systems~\cite{Dacrema2019AreApproaches,Rendle2020NeuralRevisited,Ludewig2019EmpiricalAlgorithms}, we believe that research on APC would benefit from a formal characterization of the requirements expected from suitable solutions for large-scale APC applications, which we~provide~in~the~remainder~of~this~paper.

\section{Large-Scale Playlist Continuation}
\label{section:continuation}

In this section, we present and analyze our proposed framework to build scalable yet effective APC models for large-scale applications. 

\subsection{Objectives and Requirements} 
\label{overview}

Our main goal is to provide a comprehensive set of guidelines for implementing APC models that are scalable by design. Additionally, we aim to maintain flexibility and generality in our approach.

Specifically, in this paper, our definition of \textit{scalability} is twofold. Firstly, we want to build models that can handle large datasets with millions of songs and users, as is often required in real-world applications on music streaming services. Furthermore, insights from industrial practitioners have emphasized that scalability is also essential during the \textit{inference} phase \cite{Rendle2020NeuralRevisited}.
In practice, while regularly training an APC system is necessary to incorporate new data (e.g., get information on new songs and playlists, and consequently update model parameters), these updates can be purposely delayed until a batch of new events has been observed. This allows model training operations to be performed \textit{offline on a regular schedule}, such as once per day, and enables the model to take more time to update its parameters \cite{2011RecommenderHandbook}. In contrast, the inference process, repeatedly making recommendations to users, must~be~accomplished:
\begin{itemize}
    \item \textit{Online} on the service: as users regularly create and update playlists, the exact song sequence that must be extended by the APC model can \textit{not} be processed offline in advance.
    \item With minimal delay: to minimize latency costs when providing updated recommendations to users. For example, on a global service like Deezer, an inference time longer than \textit{a few tens of milliseconds} would be unacceptable in production. 
\end{itemize}

For these reasons, scalability of inference will be particularly crucial in our framework. Consequently, we will aim to minimize the number of \textit{online} operations whose complexity depends on the size of the dataset, described by $N$, $L$, $M$, and $\text{max}_{1 \leq k \leq M}|\mathcal{M}_k|$. In practice, the number of songs $N$ is the largest of these parameters. Therefore, we will prioritize the \textit{offline} pre-computation of operations depending on $N$, whenever possible. 



\subsection{A Scalable Framework for APC}
\label{section:framework}

We now formally introduce our framework for APC at scale. 
We focus\footnote{\label{focus}This focus is made without loss of generality. Indeed, single-stage APC models stemming from our scalable framework could be integrated into two-stage/ensemble systems such as the ones from Section~\ref{s223}. Also, while embedding representations are ubiquitous in the related work from Sections~\ref{s221} and \ref{s222}, our framework could be adapted to settings where $\mathbf{h}_p$ and $\mathbf{h}_s$ capture more general descriptive~information.} on single-stage models $f \colon \mathcal{P} \times \mathcal{S} \to \mathbb{R}$ directly scoring each song $s \in \mathcal{S}$ with a similarity score $f(p,s)$ for extending some playlist $p~\in~\mathcal{P}$. Also, we focus\textsuperscript{\ref{focus}} on APC models learning ``embedding'' vectorial representations for any playlist $p$ and song $s$. We denote them by $\mathbf{h}_p \in \mathbb{R}^D$ and $\mathbf{h}_s \in \mathbb{R}^D$, respectively, for some embedding dimension $D \in \mathbb{N}^{*}$. Such models can be written~as~follows:
\begin{equation}
\label{eq1}
f(p,s) = \text{SIM}(\mathbf{h}_p,\mathbf{h}_s), 
\end{equation}
for some similarity function $\text{SIM} \colon \mathbb{R}^D \times \mathbb{R}^D \to \mathbb{R}$. For each element of Equation~\eqref{eq1}, we establish properties that must be verified in our framework, following the principles from Section~\ref{overview}.

\subsubsection{Song Representation}
\label{songs_representation}

Firstly, we note that each song embedding vector $\mathbf{h}_s$ depends on the song $s \in \mathcal{S}$ and, possibly, the metadata $m_s \in \mathcal{M}$. As an embedding vector must be computed for each of the $N$ songs of the catalog, we require to perform this operation \textit{offline in advance}, with its result stored to be rapidly accessible with a simple read operation.
Consequently, song embedding vectors should \textit{not} depend on the playlist $p$ to extend. Formally, in our framework, each $\mathbf{h}_s$ vector should be computed~via~a~function\footnote{We imply the metadata in dependencies, as each $m_s$ exclusively depends on $s$.}:
\begin{equation}
\phi \colon \mathcal{S} \to \mathbb{R}^D,
\end{equation}
learning a unique vectorial representation for each song.

\subsubsection{Playlist Representation}
 On the other hand, the representation $\mathbf{h}_p$ of each playlist $p \in \mathcal{P}$ can not be computed offline in advance, as playlists can be created and updated at any time by users (see Sections~\ref{section:s21}~and~\ref{overview}). Hence, according to our requirements, the playlist-level representation learning operations should not depend on a large set of inputs. We propose to learn $\mathbf{h}_p$ solely from the song sequence $(s_{p1}, ..., s_{pl})$ characterizing $p$. Since large-scale APC usually implies dealing with sparse data~\cite{Chen2018RecSysContinuation}, our framework also aims to benefit from parameter sharing whenever possible, by adding two constraints on the function~learning~$\mathbf{h}_p$:
\begin{itemize}
    \item It should leverage song representations $\mathbf{h}_s$ as input.
    \item It should be independent of the playlist length $l \in \{1, \dots, L\}$. 
\end{itemize}
Hence, by defining $\mathcal{E} = \bigcup_{l=1}^{L} \mathbb{R}^{D \times l}$, each $\mathbf{h}_p$ should be~computed~via:
\begin{equation}
g \colon \mathcal{E} \to \mathbb{R}^D,
\end{equation} a function \textit{aggregating} representations from the songs present in the playlist~$p$ into a single playlist embedding representation $\mathbf{h}_p$.

\subsubsection{Similarity}

The final scoring operation $f(p,s) = \text{SIM}(\mathbf{h}_p,\mathbf{h}_s)$, estimating how likely each song $s$ will fit well as a continuation of some playlist $p$ using the above $\mathbf{h}_s$ and $\mathbf{h}_p$, constitutes the complexity bottleneck of the APC prediction process. Indeed, it depends on $N$, assuming that the APC model actually considers all songs from $\mathcal{S}$ when making recommendations (and not some approximate subset). Moreover, it can not be pre-computed offline beforehand, as it depends on the playlist representation, which is itself computed online when we model observes the exact song sequence to extend. Although similarity functions based on neural networks have been proposed \cite{He2017NeuralFiltering}, Rendle et~al.~\cite{Rendle2020NeuralRevisited} have demonstrated that they do not outperform the traditional inner product, whose ability to rapidly score millions of items makes it a natural choice for use as a similarity function in~our~framework: $ f(p,s) = \langle \mathbf{h}_p,\mathbf{h}_s \rangle$.
   

\subsubsection{Represent-Then-Aggregate}

\begin{figure}[t]
    \centering
    \includegraphics[width=0.85\linewidth]{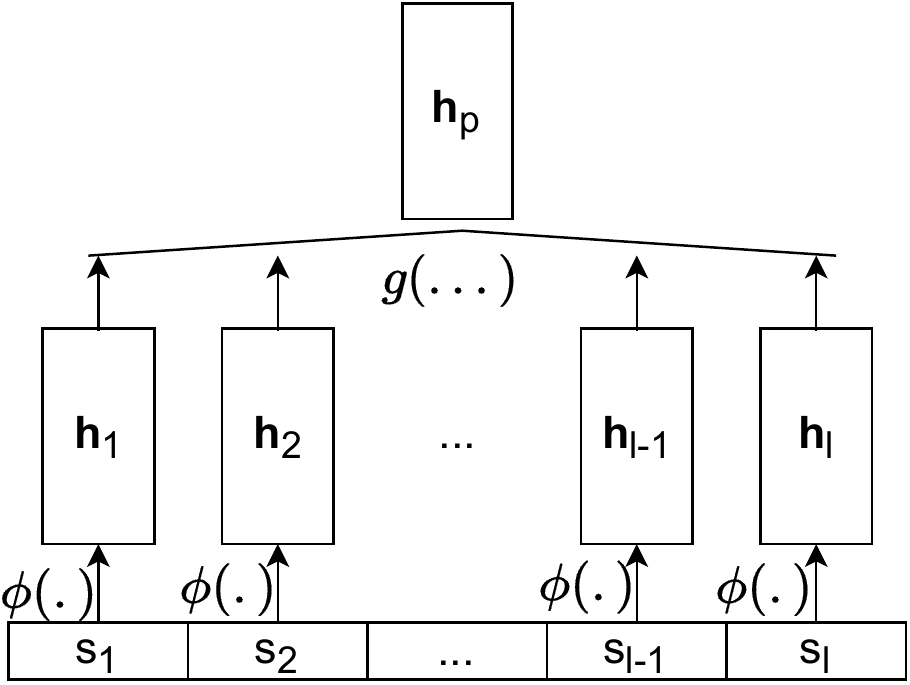}
    \caption{Our \textit{Represent-Then-Aggregate} framework for APC.}
    \label{architecture}
\end{figure}

In summary, as illustrated in Figure~\ref{architecture}, our framework constraints APC models to adopt~the~structure:
\begin{equation}
\label{eq3}
f(p,s) = \langle g(\mathbf{h}_{s_{p1}},...,\mathbf{h}_{s_{pl}}),\mathbf{h}_s\rangle, \hspace{0.1cm} \text{with } \hspace{0.1cm} \forall k \in \mathcal{S}, \mathbf{h}_k = \phi(k).
\end{equation}
As it involves a function $\phi$ representing songs and a function $g$ aggregating these representations, we refer to our framework as \textit{Represent-Then-Aggregate} (RTA) in the following.
Our experiments from Sections~\ref{section:offline_experiments}~and~\ref{section:online_experiments} will empirically confirm the scalability of various APC models complying with this framework.

\subsection{Examples of RTA Architectures}
\label{section:examples}

While ensuring scalability by design, our RTA framework remains flexible enough to incorporate a wide range of modern models.
In particular, our experiments will consider the following options for song representation learning and playlist-level aggregation.
\subsubsection{Song Representation Function $\phi$}
\begin{itemize}
    \item $\mathbf{h}_s = \phi_{e}(s) = \mathbf{e}_s$: we directly associate each song $s$ with a vector $\mathbf{e}_s \in \mathbb{R}^D$ using a representation learning model trained from playlist-song occurrence data (see~Section~\ref{rta_models}).
    \item $\mathbf{h}_s = \phi_{\text{FM}}(s) = \sum_{m \in m_s} \mathbf{e}_{m}$: we represent each song by the sum or average of embedding vectors associated~with~its~metadata, which we themselves learn using a model trained from occurrence data. We refer to this approach as FM due to its connection to Factorization Machines from Section~\ref{s221}.
    \item $\mathbf{h}_s = \phi_{\text{NN}}(s) = \text{NN}(\{\mathbf{e}_m, \forall m \in m_s\})$: we represent each song by the output of a neural network $\text{NN}$ processing song metadata. Our experiments will consider an attention-based neural architecture similar to the one of Song~et~al.~\cite{Song2018AutoInt:Networks}.
\end{itemize}
\subsubsection{Playlist Sequence Modeling / Aggregation Function $g$}
\begin{itemize}
    \item $g_{\text{AVG}}(\mathbf{h}_{s_{p_1}},...,\mathbf{h}_{s_{p_l}}) = \frac{1}{l} \sum_{i = 1}^{l}\mathbf{h}_{s_{p_i}}$: the playlist representation is the average representation of the songs it contains.
    \item $g_{\text{CNN}}(\mathbf{h}_{s_{p_1}},...,\mathbf{h}_{s_{p_l}}) = \text{CNN}([\mathbf{h}_{s_{p_1}};...;\mathbf{h}_{s_{p_l}}])$: a gated convolutional neural network (CNN) processing a concatenation of song representations learns the playlist representation. A key component of the ensemble model that won the RecSys 2018 APC challenge~\cite{Volkovs2018Two-stageScale}, the use of such gated CNN for APC was inspired by its successful use in language modeling~\cite{Dauphin2016LanguageNetworks}.
    \item $g_{\text{GRU}}(\mathbf{h}_{s_{p_1}},...,\mathbf{h}_{s_{p_l}}) = \text{GRU}([\mathbf{h}_{s_{p_1}};...;\mathbf{h}_{s_{p_l}}])$: a recurrent neural network (RNN) composed of gated recurrent units (GRU) learns the playlist representation. Our experiments will consider an architecture similar to the one of Hidasi~et~al~\cite{Hidasi2016Session-basedNetworks} but with a different loss, presented in Section~\ref{training_rta}.
    \item $g_{\text{Transformer}}(\mathbf{h}_{s_{p_1}},...,\mathbf{h}_{s_{p_l}}) = \text{Decoder}([\mathbf{h}_{s_{p_1}};...;\mathbf{h}_{s_{p_l}}])$: the Decoder part of a Transformer network~\cite{Vaswani2017AttentionNeed} learns the playlist representation. The choice of keeping only the Decoder for APC was motivated by the recent surge of GPT models, which have demonstrated state-of-the-art performances on the analogous task of sentence continuation~\cite{AlecRadford2018ImprovingPre-Training}.
\end{itemize}

\subsubsection{Limitations}
These examples illustrate the versatility of our RTA framework. Simultaneously, we are also aware of some limitations it imposes. In particular, Equation~\eqref{eq3} can not express APC architectures inspired by word2vec~models~\cite{Vasile2016Meta-Prod2VecRecommendation,Wang2016LearningRecommendation}. Indeed, they leverage \textit{different} song representations depending on whether they are used to characterize the context of other songs, or evaluated as candidates for APC. Leveraging different representations is relevant for language modeling, since words within a sentence are rarely synonyms and should not be given close representations, yet models attempt to represent words by their context, motivating the need for different representations for each word (see the discussion by Goldberg et al.~\cite{Goldberg2014Word2vecMethod}).
On the contrary, in the case of APC, we believe songs appearing in the same playlist should have close representations, i.e., songs should be similar to their context (an assumption at the core of matrix factorization techniques~\cite{koren2009matrix}). For this reason, our framework excludes these approaches in favor of the scalability induced by more~parameter~sharing.

\subsection{Example of a Training Procedure}
\label{training_rta}
While our RTA framework remains general, we suggest the following training procedure, which can be applied to optimize all models stemming from this framework. Our experiments in this paper will adopt this same procedure to train all RTA-based APC models.

Our preliminary experiments have shown that leveraging pre-trained embedding vectors from a weighted regularized matrix factorization as initial song representations improves performances. Therefore, we propose to start by factorizing
the playlist-song occurrence matrix of the APC problem under consideration, associating each song $s$ with some $\mathbf{e}_s \in \mathbb{R}^D$ used for initialization. Simultaneously, we initialize embedding vectors of metadata information by averaging vectors $\mathbf{e}_s$ of songs sharing the~same~metadata~value.

Starting from these initial representations, we jointly optimize weights of the models selected as $\phi$ and $g$, using batches of playlists of varying lengths. For each playlist $p$ of length $l~\in~\{2,\dots,L\}$, we create $l-1$ sub-playlists denoted $p_{:i}$ using the first $i$ songs of $p$, with $i~\in~\{1,\dots,l-1\}$. Then, we sample a song set $\mathcal{S}^-(p)$ from $\mathcal{S}\setminus p$. As we expect the APC model to return a high similarity scores for $f(p_{:i},p_{s_{i+1}})$ and a lower score for songs that are absent from $p$, our procedure optimizes model weights via
gradient descent minimization~\cite{goodfellow2016deep} of the loss $\mathcal{L}(p) = \mathcal{L}_{\text{pos}}(p) + \mathcal{L}_{\text{neg}}(p)$, with:
\begin{equation}
\label{eq8}
    \mathcal{L}_{\text{pos}}(p) = - \sum_{i=1}^{l-1}\text{log}\Big(\frac{1}{1+e^{-f(p_{:i},p_{s_{i+1}})}}\Big),
\end{equation}
and:
\begin{equation}
\label{eq9}
    \mathcal{L}_{\text{neg}}(p) = - \sum_{i=1}^{l-1}\sum_{s^-\in \mathcal{S}^-(p)}\text{log}\Big(1-\frac{1}{1+e^{-f(p_{:i},s^-)}}\Big).
\end{equation}

\section{Offline Experiments}
\label{section:offline_experiments}

We now present an experimental evaluation of our framework on the Million Playlist Dataset (MPD)~\cite{Chen2018RecSysContinuation,Zamani2019AnContinuation}. We release our source code on GitHub\footnote{\label{sourcecode} \href{https://github.com/deezer/APC-RTA}{https://github.com/deezer/APC-RTA}}, to ensure the reproducibility of our results and to encourage the future usage of our framework.

 \subsection{Experimental Setting}
 \subsubsection{Dataset} 
 We consider the entire MPD for our experiments. This dataset includes one million playlists created by Spotify users, using more than two million songs (see Table~\ref{table:datasets}). 
For each song $s$, we have metadata information\footnote{The MPD also includes the \textit{title} of each playlist as an additional information~\cite{Chen2018RecSysContinuation}. However, Zamani et~al~\cite{Zamani2019AnContinuation} have reported that this information did not significantly improve performances during the RecSys 2018 APC Challenge. Moreover, titles are often unavailable for APC, as well as for closely related tasks such as radio generation/personalization~\cite{bontempelli2022flow}. For these reasons, we omit playlist titles in these experiments.} corresponding to its artist ($\text{art}_s$), the album it belongs to ($\text{alb}_s$), the duration of the song ($\text{dur}_s$), and the number of times it occurs in the dataset, a metric referred to as popularity ($\text{pop}_s)$. Hence, $m_s = (\text{art}_s, \text{alb}_s, \text{dur}_s, \text{pop}_s)$. As $\text{dur}_s$ and $\text{pop}_s$ are numerical values that can not be easily mapped to embedding vectors, we group similar values together~into~buckets:
\begin{itemize}
 \item For the duration, we create linear buckets of 30 seconds each, with songs longer than 20 minutes being assigned to the same bucket. We have: $\text{bucket}_{\text{dur}}(s) = \text{min}(40,\left\lceil\dfrac{\text{dur}_s}{30}\right\rceil)$.
 \item For the popularity, we create buckets using a logarithmic scale, so that we distinguish smaller values while higher values are assigned to the same bucket. By setting $\alpha=$ 45~000, i.e., the highest number of occurrences observed in the dataset, we have:
 $\text{bucket}_{\text{pop}}(s) = \text{min}(100, 1 + 100 \times \left\lfloor\dfrac{\text{log}(\text{pop}_s)}{\text{log}(\alpha)}\right\rfloor)$.
\end{itemize}


\subsubsection{Task}
We consider a large-scale APC evaluation task similar to the one presented in Section~\ref{s212}. Specifically, we start by randomly sampling
20~000 playlists
of length $l \geq$ 20 songs from the MPD, to constitute a validation set and a test set of 10~000 playlists each. The remaining 980~000 playlists constitute our training set.

Then, we mask the last songs of test playlists, so that only their $n_{\text{seed}}$ first songs are visible. We consider ten different configurations, with $n_{\text{seed}}$ varying from 1 to 10, randomly selecting 1~000 test playlists for each configuration. Our experiments consist in assessing the ability of several APC models trained on the 980~000 train playlists (see Section~\ref{s42}) to retrieve the masked songs of each test playlist. 
Consistently with the RecSys 2018 APC Challenge, we require each APC model to predict a ranked list of $n_{\text{reco}} = 500$ candidate songs to continue each test playlist.

\subsubsection{Metrics}

In the following, we analyze nine different metrics to evaluate each model on the specified APC task. Firstly, we consider five \textit{accuracy-oriented} metrics. Besides the prevalent \text{Precision} and \text{Recall} scores~\cite{Schedl2018CurrentResearch}, we report the three metrics used for evaluation during the RecSys~2018~APC~Challenge~\cite{Chen2018RecSysContinuation}:
 \begin{itemize}
 \item \text{Normalized Discounted Cumulative Gain (NDCG)}: acts as a measure of ranking quality. It increases when the ground truth masked songs are placed higher in the ranked list of candidate songs recommended by the APC model~\cite{wang2013theoretical}.
 \item \text{R-Precision}: jointly captures the proportion of masked songs and artists recommended by the model. Artist matches increase the score even if the predicted song is incorrect~\cite{Chen2018RecSysContinuation}.
     \item \text{Clicks}: indicates how many batches of ten candidate songs must be recommended (starting with the top ten candidates) before encountering the first ground truth song. Unlike previous metrics, Clicks should, therefore, be~minimized~\cite{Chen2018RecSysContinuation}.
 \end{itemize}
Zamani et al.~\cite{Zamani2019AnContinuation} provide exact formulas for these three metrics.
In addition, we compute two \textit{popularity-oriented} scores, described by Ludewig and Jannach~\cite{Ludewig2018EvaluationAlgorithms}. They monitor the tendency of each model to cover the entire catalog when providing recommendations:
 \begin{itemize}
 \item \text{Coverage}: computes the percentage of songs from the catalog recommended at least once during~the~test~phase.
 \item \text{Popularity bias}: averages the popularity of recommended songs, computed by counting the occurrences of each song in the training set and applying min-max normalization~\cite{Ludewig2018EvaluationAlgorithms}. Low scores indicate that less popular songs~are~recommended.
 \end{itemize}
Finally, to compare the \textit{scalability} of each model, we compute:
 \begin{itemize}
 \item \text{Training time}: the time required for a model to learn from the training set until no more improvement on the validation set could be observed, regarding any accuracy-oriented~metrics.
 \item \text{Inference time}: the average time required to recommend a list of $n_{\text{reco}} = 500$ candidate songs to extend a test playlist.
 \end{itemize}


\begin{table*}[t]\centering
\caption{Automatic Playlist Continuation (APC) on the Million Playlist Dataset~(MPD)~\cite{Chen2018RecSysContinuation}, using various models stemming from our proposed Represent-Then-Aggregate (RTA) framework and other baselines. Scores are computed on test playlists and averaged for $n_{\text{seed}}$ varying from 1 to 10. All models are required to provide a ranked list of $n_{\text{reco}} = 500$ candidate songs to continue each test playlist.
All embedding models learn representations of dimension $D = 128$, with other hyperparameters set as described in Section~\ref{s42}. The two columns \textit{``Acceptable for MSS?''} indicate whether the reported training and inference times would be acceptable ($\checkmark$) or not ($\times$) for APC on a Music Streaming Service (MSS), and are discussed in Section~\ref{s432}.}
\label{table:results}

\resizebox{1.0\textwidth}{!}{
\begin{tabular}{r||ccccc|cc|cc|cc}
\toprule
\textbf{Models} &\textbf{Precision} &\textbf{Recall}&\textbf{R-Precision} &\textbf{NDCG}&\textbf{Clicks} &\textbf{Popularity} &\textbf{Coverage} &\textbf{Total time for}& \textbf{Acceptable} & \textbf{Inference time} & \textbf{Acceptable}\\
&\textbf{(in \%)} &\textbf{(in \%)}&\textbf{(in \%)} &\textbf{(in \%)}& \textbf{(in number)}&\textbf{(in \%)}&\textbf{(in \%)} & \textbf{model training} & \textbf{for MSS?} & \textbf{by test playlist}  & \textbf{for MSS?}\\
\midrule
\midrule 
\underline{Baselines} & & & & & & & & & & & \\
SKNN & $4.93\pm{0.09}$& $36.94\pm{0.49}$& ${21.42\pm{0.30}}$& ${27.66\pm{0.40}}$& $3.82\pm{0.22}$& $14.86\pm{0.17}$& $12.64$ & 4~min & \textbf{\checkmark} & $\sim$~0.5~sec & $\times$ \\ 
VSKNN & $4.93\pm{0.09}$& $36.83\pm{0.49}$& ${21.27\pm{0.30}}$& ${27.54\pm{0.40}}$& $3.84\pm{0.22}$& $15.18\pm{0.17}$& $12.28$ &4~min & $\checkmark$  & $\sim$~0.5~sec & $\times$\\ 
STAN & $4.32\pm{0.09}$& $32.73\pm{0.48}$& $19.12\pm{0.28}$& $24.26\pm{0.38}$& $4.73\pm{0.23}$& $13.43\pm{0.16}$& $\boldsymbol{27.03}$ & 4~min & $\checkmark$ & $\sim$~0.5~sec & $\times$\\ 
VSTAN & $4.59\pm{0.09}$& $34.84\pm{0.49}$& $20.33\pm{0.30}$& $25.99\pm{0.40}$& $4.49\pm{0.23}$& ${11.35\pm{0.16}}$& ${20.54}$& 3~min & $\checkmark$ & $\sim$~0.5~sec & $\times$ \\
\midrule
\underline{RTA Models} & & & & & & & & & & & \\
MF-AVG & $4.43\pm{0.09}$& $32.64\pm{0.46}$& $19.80\pm{0.28}$& $24.46\pm{0.37}$& $5.95\pm{0.28}$& $21.37\pm{0.19}$& $1.07$ & 15~min & $\checkmark$ & <~0.01~sec & $\checkmark$\\ 
MF-CNN & $5.00\pm{0.09}$& $37.89\pm{0.45}$& $21.15\pm{0.26}$& $26.83\pm{0.35}$& $3.13\pm{0.19}$& $15.43\pm{0.14}$& $3.99$ &$\sim$~5~h & $\checkmark$ & <~0.01~sec & $\checkmark$\\ 
MF-GRU & $5.16\pm{0.09}$& $39.16\pm{0.46}$& $21.83\pm{0.27}$& $28.22\pm{0.36}$& $2.77\pm{0.18}$& $12.70\pm{0.13}$& $3.18$ & $\sim$~7~h & $\checkmark$  & <~0.01~sec & $\checkmark$\\ 
MF-Transformer & ${5.20\pm{0.09}}$& ${39.76\pm{0.47}}$& $\boldsymbol{{22.30\pm{0.28}}}$& ${29.04\pm{0.38}}$& $2.60\pm{0.17}$& $10.46\pm{0.13}$& $5.35$ & $\sim$~5~h& $\checkmark$ & <~0.01~sec & $\checkmark$\\ 
FM-Transformer & $\boldsymbol{5.31\pm{0.09}}$& $\boldsymbol{40.46\pm{0.47}}$& $22.29\pm{0.27}$& $\boldsymbol{29.21\pm{0.37}}$& $2.52\pm{0.17}$& $11.74\pm{0.13}$& $10.18$ & $\sim$~6~h & $\checkmark$ & <~0.01~sec & $\checkmark$\\ 
NN-Transformer & ${5.26\pm{0.09}}$& ${40.17\pm{0.47}}$& ${22.18\pm{0.27}}$& ${29.14\pm{0.37}}$& $\boldsymbol{2.17\pm{0.15}}$& $\boldsymbol{10.33\pm{0.13}}$& $11.81$ & $\sim$~6~h & $\checkmark$  & <~0.01~sec & $\checkmark$ \\ 

\bottomrule
\end{tabular}
\label{sequenceevaluation}
}
\end{table*}

\subsection{APC Models}
\label{s42}
We compare the scalability and performance of ten APC models.
 \subsubsection{Models based on our RTA framework}
 \label{rta_models}
Firstly, we consider six different RTA models, built by leveraging the following song representation and playlist aggregation/modeling~functions:
 \begin{itemize}
    \item MF-AVG\footnote{\label{footnoteensemble}  The ensemble system that won the RecSys~2018~Challenge included~such~a~model~\cite{Volkovs2018Two-stageScale}.}: employs the representation function $\phi_{e-\text{MF}}$ (see below), combined with the aggregation function $g_{\text{AVG}}$.
    \item MF-CNN\textsuperscript{\ref{footnoteensemble}}: $\phi_{e-\text{MF}}$ combined with $g_{\text{CNN}}$.
    \item MF-GRU: $\phi_{e-\text{MF}}$ combined with $g_{\text{GRU}}$.
    \item MF-Transformer: $\phi_{e-\text{MF}}$ combined with $g_{\text{Transformer}}$.
    \item FM-Transformer: $\phi_{\text{FM}}$ combined with $g_{\text{Transformer}}$.
    \item NN-Transformer: $\phi_{\text{NN}}$ combined with $g_{\text{Transformer}}$.
 \end{itemize}
In the above list, we use the notation from Section~\ref{section:examples}. We additionally denote by $\phi_{e-MF}$ a particular example of representation learning function $\phi_{e}$. Specifically, $\phi_{e-MF}$ initializes song embedding vectors using the weighted regularized matrix factorization mentioned in Section~\ref{training_rta} and directly refines these representations via the minimization of the loss defined in this same section. All six models learn embedding vectors of dimension $D = 128$. We optimized them via the procedure of Section~\ref{training_rta}, using stochastic gradient descent~\cite{goodfellow2016deep} with batches of 128 playlists and 100 negative samples. We tuned all hyperparameters to maximize NDCG scores on the validation set, using the \textit{Optuna} library~\cite{Akiba2019Optuna:Framework} for efficient hyperparameter search. For brevity, we report all optimal hyperparameter values in our GitHub repository\textsuperscript{\ref{sourcecode}}. All $g_{\text{CNN}}$, $g_{\text{GRU}}$, $g_{\text{Transformer}}$, and $\phi_{\text{NN}}$ models had from 1 to 3 layers. The width of hidden layers went from 128 to 1024 units. The kernel size for $g_{\text{CNN}}$ models went from 2~to~5. The number of heads for $g_{\text{Transformer}}$ and $\phi_{\text{NN}}$ was a power of 2 ranging from $2^1$ to $2^6$. We tested learning rates ranging from $10^{-3}$ to 1, weight decays from $10^{-9}$ to $10^{-4}$, and dropout rates from 0 to 0.5~\cite{srivastava2014dropout}. For every model, we halved the learning rate at each epoch and performed early stopping using the validation~set~\cite{goodfellow2016deep}.
We used Python, training models on a single CPU machine with 25~GB of RAM and a Tesla~P100~GPU~accelerator.

\subsubsection{Baselines}
By studying these six different RTA models, our goal is to measure how well various modern methods would scale and perform using our framework, including techniques previously overlooked for large-scale APC. To establish reference  points, we also concurrently evaluate and compare four non-RTA~baselines:
\begin{itemize}
    \item SKNN~\cite{Hariri2015Adapting}: a session-based nearest neighbors approach that, despite its apparent simplicity, can reach competitive performances with respect to CNN and GRU models~\cite{Ludewig2019EmpiricalAlgorithms}.
    \item VSKNN~\cite{Ludewig2018EvaluationAlgorithms}: a variant of SKNN considering the song order within the playlist sequence as well as its popularity.
    \item STAN~\cite{Garg2019SequenceRecommendations}: a variant of SKNN capturing the position of the song in the playlist, information from past playlists, and the position of recommendable songs in neighboring playlists.
    \item VSTAN~\cite{Ludewig2019EmpiricalAlgorithms}: combines the ideas of STAN and VSKNN into a single approach, and adds sequence-based~item~scoring.
\end{itemize}
We selected these four baselines for two reasons:
\begin{itemize}
    \item Firstly, regarding performances, they tend to surpass alternatives in the recent empirical analysis of Ludewig~et~al.~\cite{Ludewig2019EmpiricalAlgorithms}.
    \item Secondly, although they were evaluated on smaller datasets, we could train them on the MPD using~our~machines.
\end{itemize}
 We tuned all hyperparameters by maximizing NDCG validation scores. Our set of  possible values for hyperparameters is similar to Ludewig~et~al.~\cite{Ludewig2019EmpiricalAlgorithms}. These four models require timestamps of actions performed on the items of each sequence. As the MPD includes the timestamp of the latest update of each playlist and the duration of each song, we made the simplifying assumption that each song had been added right after the previous one had~been~listened~to~once. 

\subsubsection{Scope Limitation}

For completeness, we note that we initially considered other baselines, which we ultimately discarded for scalability reasons. Most notably, we will not compare to the multimodal CF model from Yang~et~al.~\cite{Yang2018MMCF:Continuation} and the hybrid model from Rubtsov~et~al.~\cite{Rubtsov2018AContinuation}. While reaching promising results during the RecSys 2018 APC Challenge, both of these two models require more than \textit{four full days} to train on the MPD using their respective implementations (even on a machine with 28 cores and 200GB of RAM, for the latter one~\cite{Rubtsov2018AContinuation}). Considering that the MPD represents only a small fraction of playlists created by Spotify users, these models would require even more time and resources to be used by such a service.
As scalability is at the core of our work, our experiments, therefore, discard these~demanding~baselines.

In addition, as explained in Section~\ref{section:framework}, we focus on single-stage APC models, directly processing and scoring all songs from $\mathcal{S}$. For clarity of exposure and comparisons, our experiments do not include systems aggregating \textit{multiple} song scoring models. Nonetheless, it bears mentioning that the ten models from our experiments could easily be used as the first ``candidate retrieving'' part of a two-stage architecture (see Section~\ref{s223}), or as components in a larger ensemble system (e.g., Volkovs~et~al.~\cite{Volkovs2018Two-stageScale} themselves combined MF-AVG and MF-CNN in their ensemble during the challenge). As future work, we plan to extend the scope of this paper, by examining the use of several RTA models \textit{in conjunction} for large-scale~APC.

\begin{figure*}[ht]
  \centering
  \includegraphics[width=0.95\textwidth]{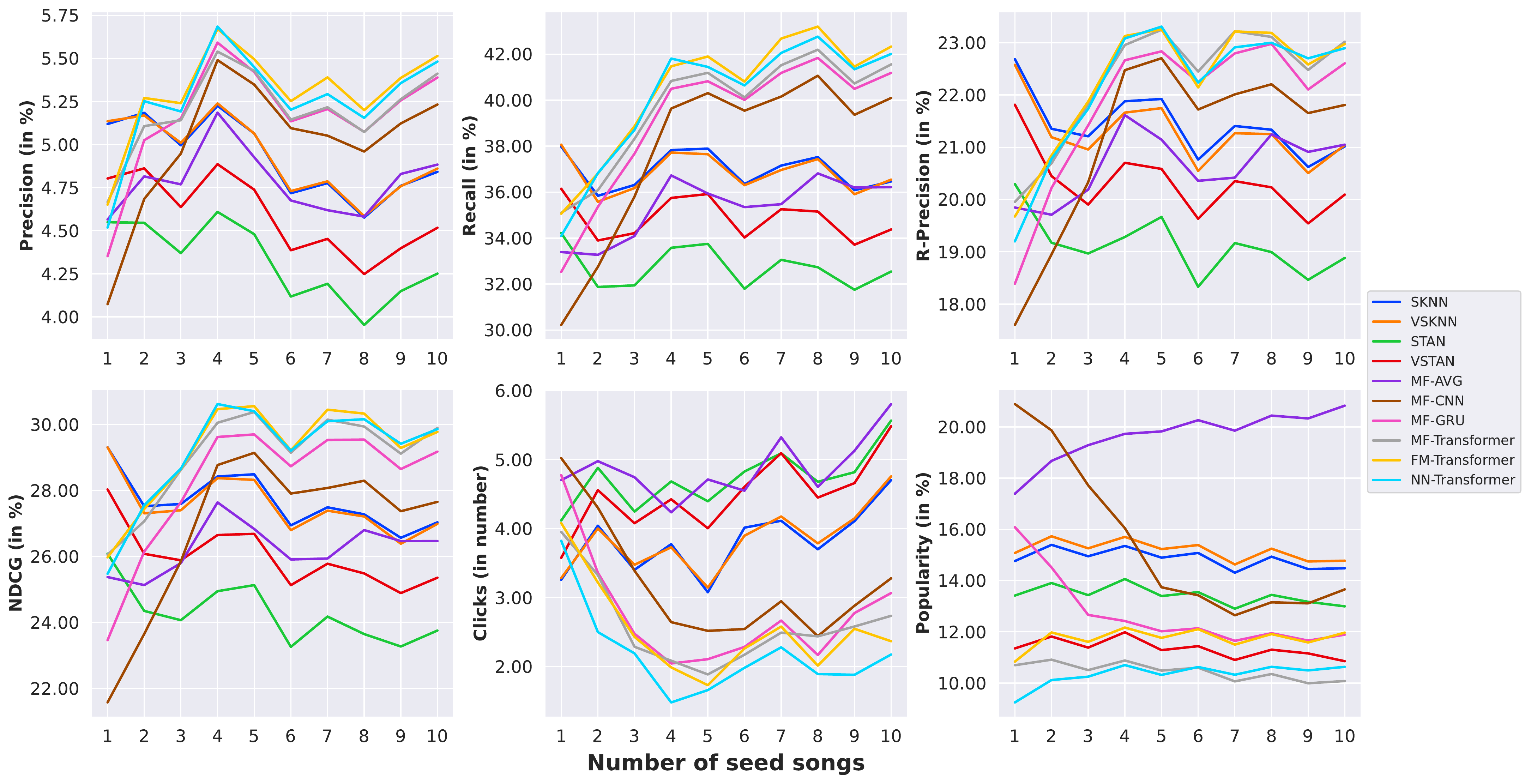}
  \caption{Automatic Playlist Continuation (APC) on the Million Playlist Dataset~(MPD)~\cite{Chen2018RecSysContinuation}, using the setting and models from Table~\ref{table:results}. Contrary to this previous table, we split scores depending on $n_{\text{seed}}$, i.e., the number of visible songs in each test playlist.}
  \label{figure:offline_results}

\label{clicks}

\end{figure*}
 \subsection{Results}
 \label{s43}
  
  \subsubsection{Performance}
  \label{s431}
Table~\ref{table:results} reports all scores\footnote{Recall: we aim to \textit{minimize} Clicks and Popularity. We report no confidence intervals for Coverage as it considers the \textit{total} number of recommended songs for the test set.} on test playlists, averaged for $n_{\text{seed}}$ varying from 1 to 10, along with 95\% confidence intervals. Firstly, we confirm previous insights~\cite{Ludewig2019EmpiricalAlgorithms} claiming that properly tuned nearest neighbors models, such as SKNN and VSKNN, can reach comparable performances with respect to some neural models (e.g., with a 27.66\% NDCG for SKNN, vs. 26.83\% for MF-CNN and 28.22\% for MF-GRU). Regarding accuracy-oriented metrics, STAN, VSTAN, and MF-AVG tend to underperform, while Transformer-based models using our RTA framework achieve the best results (e.g., with a top 40.46\% Recall for FM-Transformer). The FM-Transformer variant, leveraging Factorization Machine methods for song representation learning, slightly surpasses MF-Transformer and NN-Transformer on three metrics, even though all scores are statistically very close. We note that, in particular, Transformer-based models outperform alternatives in terms of ranking quality (NDCG, Clicks), a valuable property for real-world usage. This validates the relevance of our work, which specifically aims to facilitate and support the incorporation of such modern sequence modeling techniques into~large-scale~APC~systems.



Figure \ref{figure:offline_results} shows the evolution of different metrics as the number of visible songs $n_{\text{seed}}$ in test playlists increases. While confirming previous conclusions, the figure also highlights that sequence modeling neural architectures struggle on very short playlists. For $n_{\text{seed}} \leq 2$, VSKNN and SKNN even surpass Transformers. Another interesting observation from Figure~\ref{figure:offline_results} and Table~\ref{table:results} relates to popularity-oriented scores. 
On average, MF-AVG provides the most mainstream recommendations (with a top 21.37\% Popularity score, and a bottom 1.07\% Coverage of the catalog), followed by VSKNN and SKNN. On the contrary, the good performances of Transformers do not stem from focusing only on popular songs. Interestingly, STAN and VSTAN recommend the most diverse songs (e.g., with a top 27.03\% Coverage for STAN), but at the cost of a~deteriorating~performance.

\subsubsection{Scalability}
\label{s432} We now discuss scalability metrics. They are the ones that best showcase the relevance of our RTA framework for large-scale industrial applications. While baselines were faster to train (3 to 4 minutes on our machines), the training times associated with RTA-based models (15 minutes to a few hours) would remain acceptable for APC on a music streaming service. Indeed, as explained in Section~\ref{overview}, training operations do not  have to be executed in real-time on such a service. They can be performed \textit{offline} on a regular schedule (e.g., once per day) to update the parameters of the production model.
In practice, all models evaluated in Table~\ref{table:results} could integrate new input data overnight.

In contrast, the \textit{inference} process could not be delayed by a music streaming service. As detailed in Section~\ref{overview}, it would usually need to be \textit{repeatedly} performed \textit{online}, to provide recommendations \textit{in real-time} to many users.
In particular, we explained in this previous section that an inference time longer than a few tens of milliseconds by playlist would be unacceptable in the production environment of a service like Deezer.
In Table~\ref{table:results}, baselines require 500 milliseconds (0.5 seconds) to extend each test playlist with a list of 500 recommended songs. All RTA models perform this same operation in less than 10 milliseconds, even the most advanced models like NN-Transformer, using an attention-based network for song representation and a Transformer for playlist modeling. While all models could benefit from better hardware, overall, the inference times associated with our RTA models make them more suitable for production usage. These results highlight the ability of our framework to integrate complex but promising models, previously overlooked for large-scale APC, into effective systems meeting the scalability requirements associated with real-world~industrial~applications.


\section{Online Experiments} 
\label{section:online_experiments}


This section showcases how our framework recently helped Deezer improve APC at scale. Our objective in this section is not to re-evaluate all models from Section~\ref{section:offline_experiments}, but rather to complete our analysis by further illustrating the practical relevance of our framework, through its successful application in an~industrial~setting.


\subsection{APC on a Music Streaming Service}

\begin{figure}[t]
  \centering
  \includegraphics[width=0.65\linewidth]{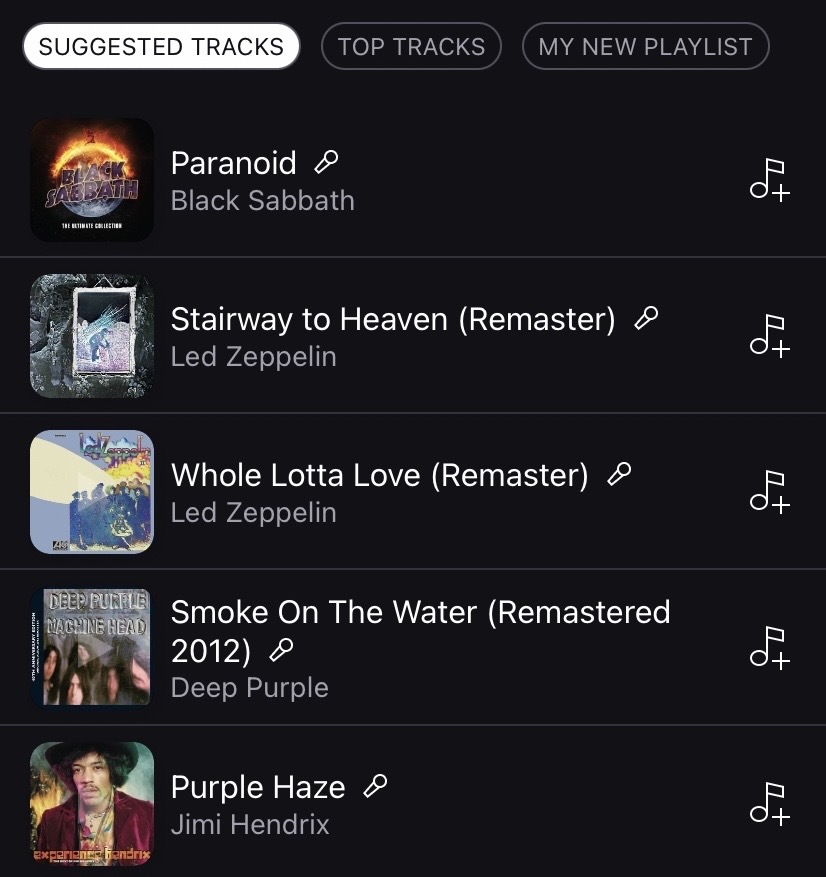}
  \caption{APC on Deezer: songs recommended by our MF-Transformer to continue the playlist from Figure~\ref{figure:playlist1}.}
\label{figure:playlist2}
\end{figure}

The global music streaming service Deezer offers an APC feature, illustrated in Figure~\ref{figure:playlist2}. It allows millions of Deezer users to automatically extend the playlists they have created on the service. In April~2022, we conducted a large-scale online A/B test on this feature, aiming to improve recommendations by leveraging the possibilities induced by our RTA framework.
Our reference system, used in production before the test and denoted ``Reference'' in the following, exploited a collaborative filtering model analogous to MF-AVG. It also incorporated various internal rules on the frequency of artist and album appearances in the list~of~recommended~songs.

During the test, we instead used our RTA-based MF-Transformer, one of the three best models from Section~\ref{section:offline_experiments}, to provide APC to a randomly selected cohort of test users.
We trained MF-Transformer and Reference on a private dataset of 25 million user playlists. Both models chose recommendations from a pool of two million recommendable songs.
As illustrated in Figure~\ref{figure:playlist2}, users requesting a playlist continuation were presented with a vertically scrollable list of 100 songs, with the five first ones being initially visible on mobile screens. Users were unaware of the algorithm change.
 
 \subsection{Online A/B Test Results}

Firstly, our RTA framework proved to be a valuable asset in integrating  MF-Transformer into Deezer's production environment. It permitted the successful integration of this model, resulting in its ability to perform APC at a minimal latency cost for users.
As an illustration, we observed a 99\up{th} percentile inference time of only 12 milliseconds on the service, unnoticeable by users (this corresponds to the time to compute the playlist representation, score two million songs, and return 100 recommendations, using an AMD EPYC 7402P CPU machine with ten threads by inference process). This example demonstrates how our framework can enable practitioners to effectively leverage modern APC models such as Transformers, that may have otherwise been left out for~scalability~reasons.

\begin{figure}[t]
  \centering
  \includegraphics[width=0.96\linewidth]{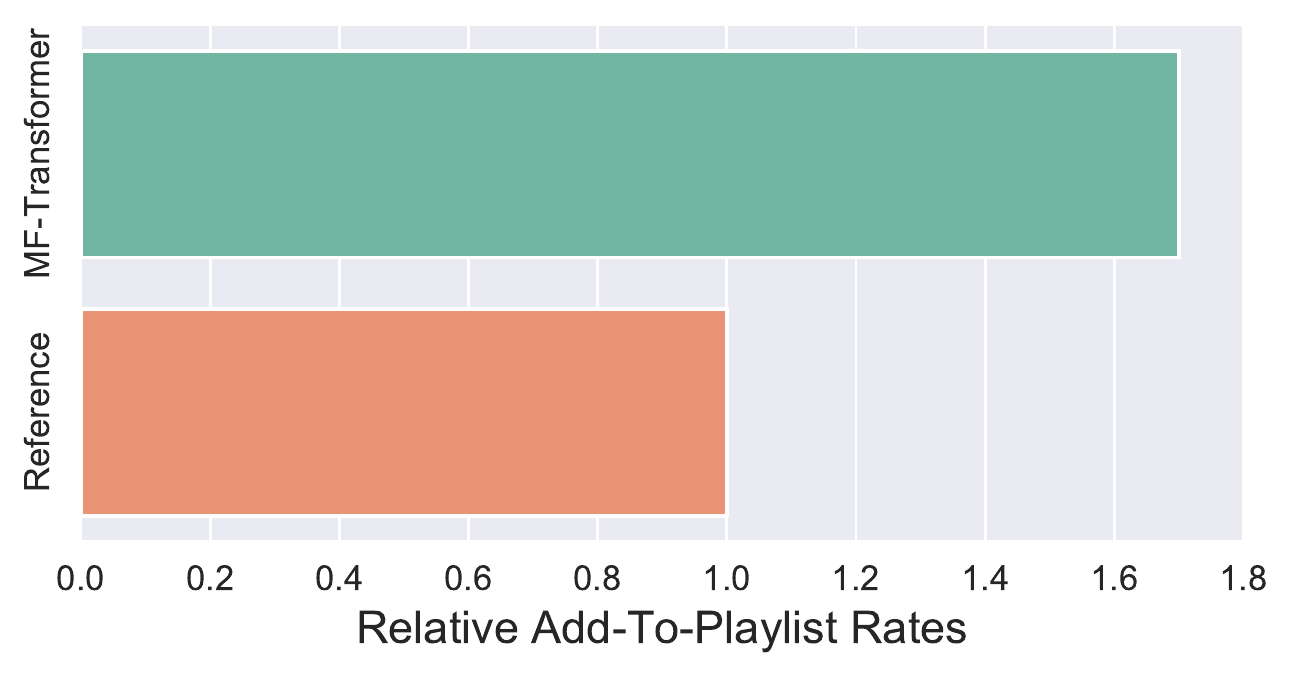}
  \caption{Online A/B test: relative Add-To-Playlist rates compared to the reference model from Deezer. Differences are statistically significant at the 1\% level (p-value < 0.01).}
\label{fig:online_results}
\end{figure}

 Regarding performances, our test confirmed the superiority of MF-Transformer over Reference on Deezer. Using MF-Transformer improved our internal performance indicators for the APC feature, such as the Add-To-Playlist rate, i.e., the percentage of recommended songs added to playlists by users.
 For confidentiality reasons, we do not report exact Add-To-Playlist rates in this paper, nor the number of users involved in each cohort. Instead, Figure \ref{fig:online_results} presents \textit{relative} Add-To-Playlist rates with respect to Reference. On average, users exposed to MF-Transformer added 70\% more recommended songs to their playlists than those in the reference~cohort.



 Following this conclusive test, MF-Transformer has been deployed to all users.
Future experiments will evaluate other APC models on the service, such as FM-Transformer and NN-Transformer. Additionally, we plan to study the application of our framework to related sequential tasks on Deezer, including~radio~personalization. Lastly, as recent work~\cite{knees2022bias} identified biases in playlist datasets that can be reproduced by recommender systems, we intend to further study these important aspects in the context of~our~framework.

\section{Conclusion}
\label{section:conclusion}
In this paper, we introduced a general framework to build scalable APC models meeting the expected requirements associated with large-scale industrial applications, e.g., on music streaming services.
We provided a detailed overview of its possibilities and limitations, showing its versatility in incorporating a wide range of advanced representation learning and sequence modeling techniques, often overlooked in previous large-scale APC experiments due to their complexity. We demonstrated the empirical relevance of our framework through both offline experiments on the largest public dataset for APC, and online experiments, improving APC at scale on a global music streaming service.
Besides the already discussed future tests on this same service, our future work will consider multi-modal song representation learning models (e.g., processing audio signals and lyrics in addition to usage data) to enhance the representation part of our framework. We will also aim to incorporate the ability for it to adapt to user~feedback~in~real-time.
 
\bibliographystyle{ACM-Reference-Format}
\bibliography{references}

\end{document}